\documentclass[a4paper]{article}
\usepackage{nameref,hyperref}
\usepackage[utf8]{inputenc}
\usepackage{amssymb}
\usepackage{booktabs}
\usepackage[english]{babel}
\usepackage[utf8]{inputenc}
\usepackage{amsmath}
\usepackage{booktabs}
\usepackage{graphicx}
\usepackage[colorinlistoftodos]{todonotes}

\title{Ideological Consumerism in Colombian Elections, 2015: Links between Political Ideology, Twitter Activity and Electoral Results}

\author{Juan C. Correa$^{1\footnote{Corresponding author: (Juan C. Correa) E-mail: \url{juanc.correan@konradlorenz.edu.co}}}$ \& Jorge Camargo$^{2}$}

\date{\small $^{1}$ Facultad de Psicología. Fundación Universitaria Konrad Lorenz. \\
$^{2}$ Facultad de Matemáticas e Ingenierías. Fundación Universitaria Konrad Lorenz. Bogotá, Colombia}
	
\begin{document}
\maketitle

\begin{abstract}
Propagation of political ideologies in social networks has shown a notorious impact on voting behavior. Both the contents of the messages (the ideology) and the politicians' influence on their online audiences (their followers) have been associated with such an impact. Here we evaluate which of these factors exerted a major role in deciding electoral results of the 2015 Colombian regional elections by evaluating the linguistic similarity of political ideologies and their influence on the Twitter sphere. The electoral results proved to be strongly associated with tweets and retweets and not with the linguistic content of their ideologies or their Twitter followers. Suggestions on new ways to analyze electoral processes are finally discussed.
\end{abstract}

\section{Introduction}
\label{S:1}

Due to the power that social networks have for promoting political mobilization and participation, politicians are using these tools to communicate their ideas and change ``political consumerism'' \cite{Stolle2005}. Political consumerism consists in turning the market into a site for politics and ethics, as consumer choices reflect personal attitudes and purchases are informed by ethical or political assessment of business and government practice \cite{Micheletti2003}. An example of this occurred when the French government opposed the approval of a UN Security Council resolution that allowed the use of military force against Iraq in 2003. By that time, sales of French wines dropped in a portion of US restaurants and in some shops, sales of French cheeses were eliminated while in some food-outlets ``French fries" were rechristened as ``Freedom fries"; a trend that shows how Americans opposed  the French position, not by taking the streets to express their views, but by using their purchasing power for spoiling French exports \cite{Stolle2005}. Political consumerism is also related with the so-called ``ideological consumerism'' roughly defined as the study of the interacting psycho-socio-cultural processes when an individual or group prefers, fosters and spreads their beliefs and quotidian practices that mediate commercial exchange \cite{Arias2009}. Our aim in this work is to offer a novel analysis of Colombian elections based on the natural language use of Politicians in Twitter.

\section{Studying Ideological Consumerism About Electoral Processes in Social Networks}

The ideological consumerism about electoral processes can also be analyzed from public opinion propagated in social networks like Facebook or Twitter. For instance, in analyzing the 2010 USA congressional elections, it was observed that right-leaning Twitter users exhibited greater levels of political activity, a more tightly interconnected social structure, and a communication network topology that facilitated the rapid and broad dissemination of political information \cite{Conover2012}. Likewise, in a randomized controlled trial of political mobilization messages delivered to 61 million Facebook users during the 2010 USA congressional elections it was observed that the messages directly influenced political self-expression, information seeking and real-world voting behavior because these messages not only influenced the users who received them but also the users' friends, and friends of friends; revealing that the effect of social transmission on real-world voting was greater than the direct effect of the messages themselves, and nearly all the transmission occurred between close friends who were more likely to have a face-to-face relationship, showing that strong ties are instrumental for spreading both online and real-world behavior in social networks \cite{Bond2012}.

Political polarization also relates with ideological consumerism. For instance, during the 2011 Canadian Federal Election, it was observed that Twitter users tended to cluster around shared political views even though they eventually interacted with other users of opposing ideologies \cite{Gruzd2014}. Furthermore, it has been noticed that the online social structure of a political party was strongly related to its ideology, and the degree of connectivity across two parties grew when they were close in the ideological space of a multi-party system like the Swiss one \cite{Garcia2015}.  Needless to say that political participation in social networks also relates with voters' feelings and attitudes. \cite{Maruyama2014} observed that participants’ average feeling and recall toward political candidates did not depend on Twitter activity which, in contrast, proved to be important for vote choice, in such a way that people who actively tweeted changed their voting choice to reflect the prevailing sentiment on Twitter. According to \cite{Park2013} these results might be partially explained by the role that a Twitter opinion leader has on individuals' involvement in political processes, since political leaders can persuade their followers directly on substantive attitudes regarding policy issues, attributions regarding the leader’s qualities, and the subsequent voting behavior \cite{Minozzi2015}.
These findings also support recent results of the 2013 elections of the Italian parliament \cite{caldarelli2014}, since the activity on Twitter in terms of volume (total tweets) and change in time provided a very good proxy of the final electoral results. Summing up, voting behavior is influenced by the political messages that are spread in social networks. Yet, given the fact that such messages reflect divergent ideologies, it might be relevant knowing if there are some similarities among their contents, and if so, which of them captures more votes and why. Here, lies the importance of analyzing  natural language use in social networks when politicians are in democratic campaign and its influence on voting behavior. 

\section{Analyzing the Ideology and the Natural Language Use of Politicians in Twitter}
\label{sec2}

A convenient way to analyze the natural language use of  politicians in Twitter can be done through the so-called \textit{word count strategies} \cite{Michalke2014, Turney2010}. This process begins with the identification of tweets whose content is related with political campaigns in the preceding weeks of the elections. All these tweets have to be merged in a single document that will be decomposed in their staple linguistic components known as ``corpus'' \cite{Pak2010}. According to \cite{Sudhof2012} a corpus  vocabulary of size $n$ can represent a ``bag of words'' of document $i$ as an n-dimensional vector $v_{i}$ where each component represents one word in the document. The value of dimension $j$ of document vector $v_{i}$ depends on the importance of the word $j$ in document $i$, being the importance quantified by contrasting the term-frequency with the inverse document frequency which formalizes the idea that a rare term or word has higher information content than expected terms like prepositions, articles and pronouns. Formally, a document vector is expressed as follows,

\begin{equation}
x_{i}=(x_{1,i},x_{2,i},x_{3,i},\ldots,x_{t,i}),\label{eq_1}
\end{equation}

and 
\begin{equation}
x_{t,i}=tf_{t}\cdot log\frac{\left|D\right|}{\left|\left\{ t\in i\right\} \right|},\label{eq_2}
\end{equation}

Where $tf_{t}$ is the term frequency of the word $t$ in the document $i$, $|D|$ is the number of documents in the collection, and $log|D|/|\left\{ t\in i\right\} |$ is the inverse frequency of the documents that contain $t$. At the end of this process the so-called ``term-document matrix'' is built. This matrix returns the words that occur at least one time in any of the documents that constitute the collection of tweets of each politician. Representing the collection of tweets as a vector in the Euclidean space allows the evaluation of its linguistic similarity with a second collection of tweets, both reflecting the ideology of each politician \cite{Turney2010}. This linguistic similarity is then estimated by the Euclidean distance between document vectors \cite{Sudhof2012}. Although this procedure does not take into account the subtleties of language (like sarcasm or irony) that are commonly used in politics \cite{Bosco2013}, it offers a suitable quantitative metric for evaluating the similarity between two politicians representing different political ideologies \cite{Henry2013}. Since Colombian ideological parties have witnessed a complex series of reforms that have promoted the creation of new parties \cite{Batlle2013}, the standard approach to differentiating them consists in using the ``left-right'' spectrum with the ``Conservative Party'' and the ``Liberal Party'' at the center stage \cite{Botero2014}. We propose another classification; that of traditional parties (e.g., ``\textit{Partido Conservador Colombiano}''), independent parties (e.g., ``\textit{Movimiento Alternativo Indígena y Social}'') or alliances between them (e.g., ``\textit{Alianza entre Partido Conservador y Partido de la U}''). If political ideology plays an important role in influencing voting behavior, as it has been previously showed in other countries \cite{Palfrey1987,Chirumbolo2010,Garcia2015,Gruzd2014}, we might expect that in the case of Colombian regional elections the linguistic similarities between politicians of the same ideological affiliation should be higher than the similarities they share with politicians of other ideologies. Alternatively, given the fact that politicians can exert their influence by increasing their Twitter activity in terms of the number of followers, the number of tweets and re-tweeted messages \cite{Cha2010,Tumasjan2010,Jungherr2012}, we also might expect a statistical significant association between these metrics and their electoral results.

\section{Methodology}
\label{sec3}

Two data sets were built for the analysis. The first was composed by the official electoral results of the last democratic elections in Colombia, held in October 25, 2015. In these elections 32 new governors had to be elected across the country (one in each Colombian department). Ad hoc queries were designed and used with the  software “Import.io”  \url{https://import.io/} for retrieving the official results from the elections web page \url{http://www.colombia.com/elecciones/2015/regionales/}. In querying these results we obtained a list containing the names of political contenders, their declared political affiliation as well as the amount of received votes in their  departments. Political abstention was 39.69\%, a little bit lower than previous ones \cite{Forero2014}. The second data set was built with the ``twitteR'' package which was developed to mining short messages in this social network through the R environment \cite{Gentry2015}. This data set was composed by the tweets spread by Colombian politicians during the preceding three weeks of the elections (between October 1, 2015 and October 24, 2015). A total of 140 candidates participated in the elections and 52 of them showed an active Twitter use for promoting their own political campaign. Thus, our sample represents 37.14\% of Colombian politicians who participated in the elections for new governors. Tweets of each politician were structured as a corpus in a single text document. We then obtained 52 corpora (one for each politician) containing all the tweets that were communicated through this social network. All of these tweets were written in Spanish and given the fact that these tweets can include special characters such as HTML tags, punctuation marks, mentions (Twitter usernames preceded by the ``@'') and hashtags (thematic words preceded by the ``\#''), we removed these characters, and preserved word accents because of their relevance in Spanish language. We followed a two-step procedure for text-cleaning purposes. In the first step we applied the so-called ``stemming'' procedure, consisting in reducing inflected words to their root word. For example, the Spanish words ``\textit{gobernación}'', ``$governaciones$'', ``$gobernador$'' and ``$gobernadores$'' were reduced to the stem word ``$gobern$''. The second step consisted in removing the so-called ``stop words'' by using the stop words list included in the ``koRpus'' package \cite{Michalke2014}. These words were removed in the because of their their high frequency in natural language use \cite{Piantadosi2014} and its resulting little contribution in the linguistic representation of a text document. A total of 69,202 tweets were processed following equations \ref{eq_1} and \ref{eq_2}. Each corpus was also accompanied by the politician's username in Twitter, the amount of Twitter followers for each tweet, and the number of retweets received by each politician's tweet. The political ideology of each text document was manually assigned according to the affiliation of the politician and its membership to one of the three categories proposed at the final part of section \ref{sec2}. The linguistic similarity between pairs of Colombian political ideologies was calculated through the Euclidean distance between document vectors.

\section{Results}
\label{sec4}
\begin{figure}[h!]
  \centering
 \includegraphics[width=1\textwidth]{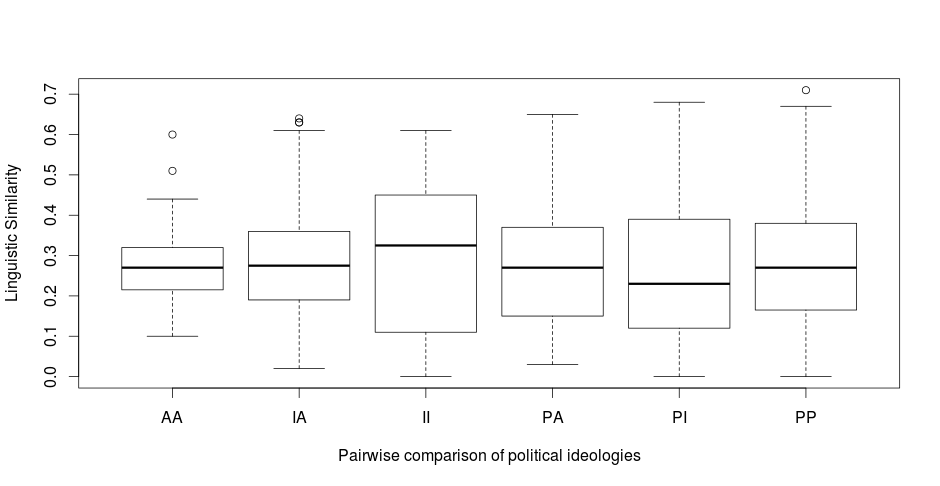}
\caption{Statistical distributions of linguistic similarity between pairs of Colombian political ideologies}
\label{A}
\end{figure}

Figure \ref{A} depicts the statistical distribution of linguistic similarity between pairs of texts (tweets-collection) representing  Colombian political ideologies; namely, the similarity between candidates both belonging to either traditional parties (PP), independent parties (II), alliances between parties (AA) or any other combination between them. The comparison reveals non-significant statistical differences (F = 1.231; df = 5; p = 0.292), since their representative candidates tend to use the same set of words for promoting theiw own electoral campaign in the preceding three weeks of the Colombian regional elections (see Table \ref{tab:table1}). 

\begin{table}[h!]
  \centering
  \caption{The 30 most frequent words (roots) found in the election corpus}
  \begin{tabular}{ccc}
    \toprule
    Word & Freq.\\
    \midrule
    gobern & 50\\
    apoy & 49\\
    vot & 49\\
    octubr & 48\\
    campa & 47\\
    graci & 45\\
    part & 45\\
    gobiern & 45\\
    comun & 44\\
    dia & 44\\
    \bottomrule
  \end{tabular}
  \begin{tabular}{ccc}
    \toprule
    Word & Freq.\\
    \midrule
    candidat & 43\\
    mejor & 43\\
    hoy & 43\\
    repald & 42\\
    invit & 42\\
    hac & 42\\
    municipi & 42\\
    segu & 41\\
    departament & 41\\
    salud & 41\\
    \bottomrule
  \end{tabular}
  \begin{tabular}{ccc}
    \toprule
    Word & Freq.\\
    \midrule
    gran & 41\\
    asi & 41\\
    propuest & 41\\
    buen & 41\\
    trabaj & 41\\
    polit & 40\\
    recib & 40\\
    gent & 40\\
    tod & 40\\
    acompa & 39\\
    \bottomrule
  \end{tabular}
\label{tab:table1}
\end{table}

\begin{figure}[h!]
  \centering
 \includegraphics[width=1\textwidth]{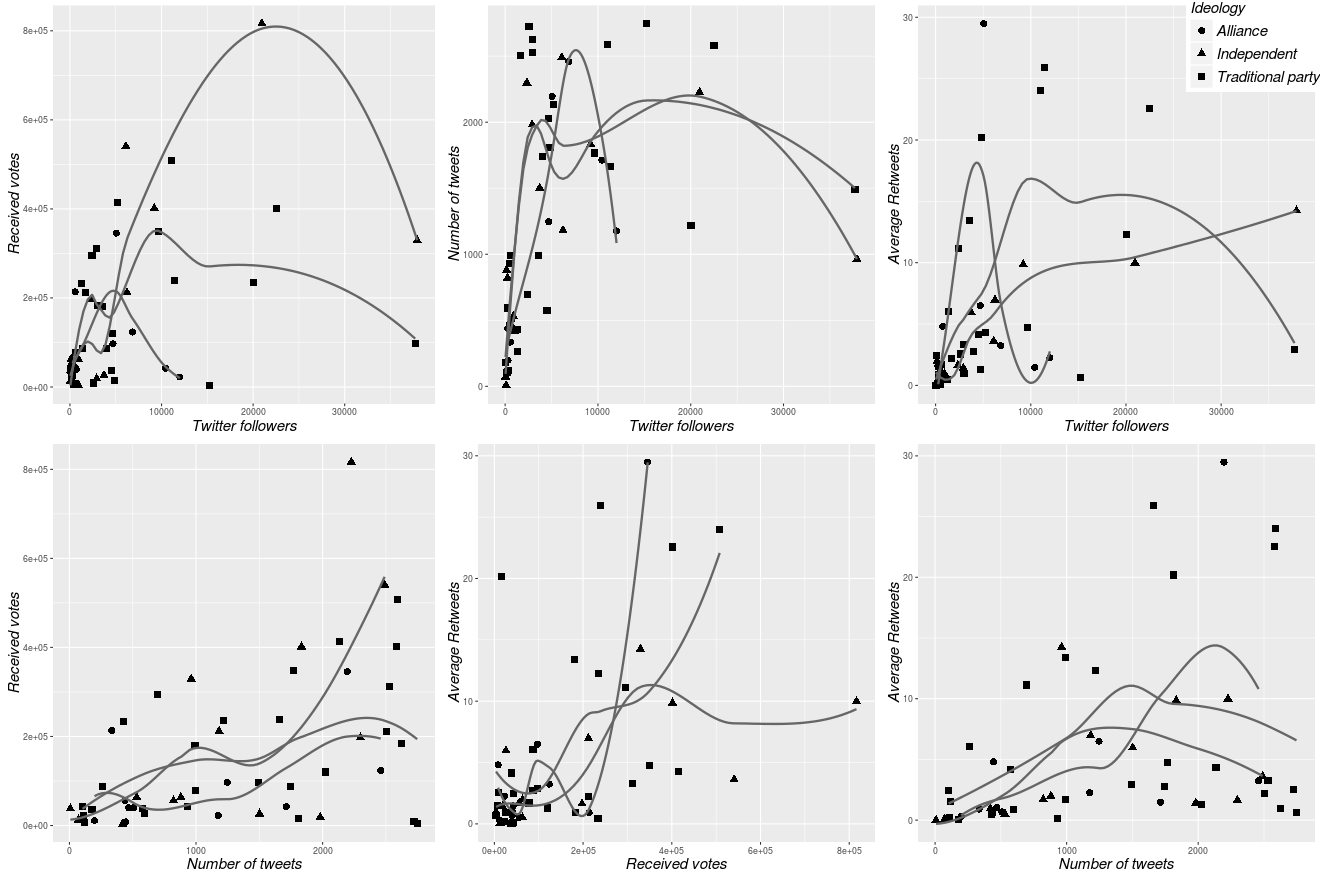}
\caption{Scatterplots of received votes and Twitter metrics by political ideology in the 2015 Colombian regional elections}
    \label{B}
\end{figure}

\newpage
Review of the Shapiro-Wilks test for normality suggests that neither the received votes (SW = 0.808; df = 52; p $<$ 0.001), nor the number of Twitter followers (SW = 0.679; df = 52; p $<$ 0.001), nor the number of tweets (SW = 0.921; df = 52; p = 0.002), nor the number of retweets (SW = 0.714; df = 52; p $<$ 0.001) showed a normal symmetric distribution. Thus, we evaluate the statistical non-parametric association between these metrics through the Spearman's rank correlation coefficient (see Figure \ref{B}).

For political alliances none of the Twitter metrics, except the one between Retweets and Twitter followers ($\rho$ = 0.682; p = 0.021), showed a significant association. In contrast, all of these metrics showed significant correlations ($\rho \geq$ 0.627; p $\leq$ 0.05) for independent parties, while for traditional parties all of these indicators proved to be statistically associated ($\rho \geq$ 0.458; p $\leq$ 0.05) except those between the number of tweets with received votes and the number of tweets with retweets. In order to clarify which of these factor exerted a major role in deciding electoral results (quantified by the received votes for each candidate) we finally test their multivariate association with a Kernel regression model \cite{Hayfield2008}. The results show an acceptable goodness of fit ($R^{2}$ = 0.787; p $\leq$ 0.001) revealing that the received votes proved to be more associated with the number of retweets (Bandwidth = 4.29; p $\leq$ 0.01)  and tweets (Bandwidth = 633.06; p = 0.09). Neither the similarity of ideological content (Bandwidth =  161148.3; p = 0.08) nor the number of followers in Twitter (Bandwidth = 49419488164; p =   0.37) proved to be important factors in relation with the received votes. 

\section{Discussion}

Our aim in this paper was to offer a different analysis of Colombian elections. We provided a convenient way to quantify the linguistic similarity of Colombian politician's tweets and use this metric for evaluating its association with received votes in the last regional elections. In addition, we employed the available metrics of Twitter activity (i.e., number of Twitter followers, tweets and retweets) to observe their association with electoral results. We showed that, at least in the Twitter sphere, Colombian politicians tend to use the same set of words, conveying a rather homogeneous message that prevents their differentiation from contenders of other ideologies. To our knowledge, this conclusion has not been found in previous studies of Colombian electoral studies \cite{Ulloa2003,Pachon2010,Moreno2011,Batlle2013,Botero2014}. Finally, but  not least, we showed that the electoral results were more associated with the  amount of retweets and tweets than with the ideological content similarity or Twitter followers. This is a contrasting result in regard to what is known about influence in Twitter \cite{Cha2010}. The number of retweets reflects the ability of a Twitter user to generate content with pass-along value. Yet, given the fact that Colombian politicians employed a quite homogeneous set of words for promoting their own political campaign, it is rather paradoxical the fact that some politicians obtain more retweets than others when they all convey a standard message in political campaign. In disentangling this paradoxical fact, future research might evaluate if the probability of retweeting a tweet has to do with the simplicity of the message instead the ideological content itself.

\section*{Acknowledgment}
We are truly indebted to Diana Onofre for her support in identifying Twitter user accounts. We also thank Professors Silvana Dakduk, Diana E. Forero and Gustavo García for comments and advice.

\bibliographystyle{ieeetr}

\label{references}
\end{document}